\documentclass{aa}  
\usepackage{graphicx}
\usepackage{txfonts}
\usepackage{amsmath,amsfonts,amssymb}
\usepackage{graphicx}
\usepackage[colorlinks=true, allcolors=blue]{hyperref}
\usepackage[capitalise]{cleveref}
\usepackage{csvsimple}
\usepackage{adjustbox}
\usepackage{rotating}
\usepackage{flafter}
\usepackage{float}

\makeatletter
\renewcommand*\aa@pageof{, page \thepage{} of \pageref*{LastPage}}
\makeatother

\begin{document} 

   \title{ {Robustness of prediction for extreme adaptive optics systems under various observing conditions} }

   \subtitle{ {An analysis using VLT/SPHERE adaptive optics data. }}

   \author{M. A. M. van Kooten
          \inst{1}\fnmsep\thanks{For author correspondence email vkooten\@strw.leidenuniv.nl}
          \and
          N. Doelman\inst{1, 2}
          \and
          M. Kenworthy
          \inst{1}
          }

   \institute{Leiden Observatory, Leiden University,
              Niels Bohrweg 2, 2333 CA Leiden\\
         \and
             TNO, Stieltjesweg 1, 2628 CK Delft}

   \date{Received October xxx, 2019; accepted xxx, 2019}

 
  \abstract
   {For high-contrast imaging (HCI) systems, such as VLT/SPHERE, the performance of the system at small angular separations is contaminated by {the wind-driven halo} in the science image. {This} halo is a result of the servo-lag error in the adaptive optics (AO) system due to the finite time between measuring the wavefront phase and applying the phase correction. One approach to mitigating the servo-lag error is predictive control.}
   {We aim to estimate and understand the {potential on-sky performance that linear data-driven prediction would provide for VLT/SPHERE under various turbulence conditions. }}
   {We used a linear minimum mean square error predictor and applied it to 27 different AO telemetry data sets from VLT/SPHERE taken over many nights under various turbulence conditions. We evaluated the performance of the predictor using residual wavefront phase variance as a performance metric.}
   {We show that prediction always results in a reduction in the temporal wavefront phase variance compared to the current VLT/SPHERE AO performance. {We find an average improvement factor of 5.1 in phase variance for prediction compared to the VLT/SPHERE residuals. When comparing to an idealised VLT/SPHERE, we find an improvement factor of 2.0.} Under our 27 different cases, we find the predictor results in a smaller spread of the residual temporal phase variance. Finally, we show there is no benefit to including spatial information in the predictor in contrast to what might have been expected from the frozen flow hypothesis. A purely temporal predictor is best suited for AO on VLT/SPHERE. }
   {Linear prediction leads to a significant reduction in phase variance for VLT/SPHERE under a variety of observing conditions and reduces the servo-lag error. Furthermore, prediction improves the reliability of the AO system performance, making it less sensitive to different conditions.  }


   \maketitle
 
%
\section{Introduction}
\label{sec:intro}  

 In the search for new exoplanets and earth analogs, dedicated high-contrast imaging (HCI) systems have allowed the {angular} separation of host stars from their surroundings to reveal cirucmstellar disks and exoplanets. The combination of extreme adaptive optics (XAO) to provide high spatial resolution, coronagraphs to suppress the host star's light, and data reduction techniques to remove residual effects,  {allows HCI systems} to reach post-processed contrasts of $10^{-6}$ at spatial separations of 200 milliarcseconds \citep{zurlo_2016}. VLT/SPHERE is a HCI system that has discovered two confirmed planets: HIP 65426b \citep{chauvin_2017} and PDS 70b \citep{Keppler_2018}. It has also discovered a vast array of debris, protoplanetary, and circumstellar disks \citep[e.g., ][]{Avenhaus_2018,sissa_2018}. Operating with a tip/tilt deformable mirror (TTDM), a 41-by-41 high order deformable mirror (HODM), and a Shack-Hartmann wavefront sensor (SHWFS) sampling at 1380 Hz, the XAO system of VLT/SPHERE, SAXO delivers Strehl ratios greater than 90 \% in the H band \citep{Beuzit_2019}. 
  
A major challenge with VLT/SPHERE (and other HCI systems to varying degrees) is the presence of the wind-driven halo (WDH) that dominates the wavefront error at small angular separations. The WDH is a manifestation of the servo-lag error and appears as a butterfly pattern in the coronagraphic/science images \citep[see][for details on the WDH]{Cantalloube_2018}. The servo-lag error is due to the finite time between the measurement of the incoming wavefront aberration (caused by atmospheric turbulence) and the subsequent applied correction.  The resulting wavefront error is due to the outdated disturbance information and the closed-loop stability constraints. Owing to this, the halo is aligned with the dominant wind direction and severely limits the contrast at small angular separations even after post-processing techniques. The servo-lag error prevents VLT/SPHERE from achieving its optimal performance when coherence times are below 5 ms \citep{Milli_2017}. 

The XAO system, SAXO, has a temporal delay of approximately 2.2 SHWFS camera frames. The HODM is controlled using an integrator with modal gain optimisation \citep{Petit_2014}. Within this framework, one solution to minimise the delay in SAXO itself is to run everything faster. However, this solution poses a number of hardware challenges with a new HODM that can run at the desired speed, a wavefront sensor camera with fast readout, and a more powerful real-time-computer.  One alternative solution is to upgrade the controller with a control scheme that predicts the evolution of the wavefront error over the time delay. In this paper, we look at the potential of prediction to improve the performance of SAXO especially when the servo-lag error is the dominant residual wavefront error source (i.e., small coherence times).
 
Many different groups have worked on predictive control as a means of improving the performance of an adaptive optics (AO) system by minimising the servo-lag error. We highlight a few results from the last 15 years.  { Prediction, within the context of optimal control, is an ingredient in finding the optimal controller. Linear quadratic Gaussian (LQG) control has been explored by ~\citet{Petit_2008} for general AO systems to perform vibration filtering with the Kalman filter. On-sky demonstrations of the LQG controller for tip-tilt/vibrational control are provided in \citet{Sivo_2014}. Laboratory work to include higher order modes for atmospheric turbulence compensation \citep{Roux_2004} has demonstrated a reduction in the temporal error showing predictive capabilities.} The H2 optimal controller (closely related to LQG) has been tested on-sky \citep{Doelman_2011} showing a reduction in the temporal error for tip-tilt control. For multi-conjugate AO systems, a temporal aspect to the phase reconstruction for each layer has been implemented;  {the spatial-angular predictor is formed by exploiting frozen flow hypothesis and making use of minimum mean square error estimator (with analytical expressions for the stochastic process) } \citep{Jackson_15}. 

{Within the HCI community, there have been many efforts to incorporate prediction into the AO control algorithm. Building on~\cite{Poyneer_06}, predictive Fourier control, proposed in~\cite{Poyneer:07}, makes use of Fourier decomposition and the closed-loop power spectral density (PSD) to find components due to frozen flow. Using Kalman filtering a predictive control law is determined resulting in a reduction of the servo-lag error. Other methods in HCI have focused on splitting the prediction
step from the controller. } This is done by first {estimating} the pseudo-open loop phase (slopes, or modes), applying a prediction filter, and then controlling the HODM using the predicted phases as input into the controller. This approach allows for a system architecture that can turn prediction on and off without affecting the control loop. A similar structure has been implemented using the CACAO real-time computer \citep{cacao_2018}. Empirical orthogonal functions (EOF) is a data-driven predictor that aims to minimise the phase variance. Implemented on SCExAO (an HCI instrument at the Subaru telescope using the CACAO real-time computer) it has been demonstrated \citep{cacao_2018} that EOF improves the standard deviation of the point-spread-function  {over a set of images. However,} the improvement is less than expected from initial simulations \citep{Guyon_2017}. Similar methods minimise the same cost function as EOF but with a different evaluation of the necessary covariance functions (see Sec.~\ref{sec:LMMSE}) as reported in \citet{vanKooten_2019} and \citet{Jensen-Clem_2019}.  {Current a posteriori tests, using AO telemetry \citep{Jensen-Clem_2019}, show an average factor of 2.6 improvement for contrast for separations from 0 to 10 $\lambda/D$. This approach to prediction will be implemented at the Keck telescope. } One benefit to separating the prediction and the control steps is that the behaviour of the input disturbance (atmospheric induced phase fluctuations) can be studied for a given system and telescope site location, along with tests performed with AO telemetry data. We take this approach in this work, building on our earlier work  \citep{vanKooten_2019}, looking solely at the predictability of the pseudo open-loop slopes under various atmospheric conditions. Previous work on prediction in the AO community has focused on one or two case(s) on-sky for demonstrations of prediction, successfully showing the feasibility of predictive control. We look at how a linear minimum mean square error (LMMSE) predictor performs a posteriori on VLT/SPHERE AO telemetry data under a large set of various observing conditions (such as guide star magnitude, coherence time, and seeing conditions). 

We organise this paper as follows: we introduce and summarise our SAXO data set in Sec.~\ref{sec:data}. In Sec~\ref{sec:methdology} we outline our methodology, including the structure of our predictor (Sec.~\ref{sec:LMMSE}). We elaborate on how we apply the predictor to the VLT/SPHERE SAXO telemetry data and we present our results in Sec.~\ref{sec:results}, discussing them in Sec.~\ref{sec:discuss}. We look at how the predictor performs under different conditions as well as the stationarity of the turbulence. The implications of the results are discussed in Sec.~\ref{sec:implications}, concluding in Sec.~\ref{sec:conclusion} including future research directions.

\section{SAXO data}
\label{sec:data}
\iftrue

\begin{figure}
\centering
\includegraphics[trim={1cm 0cm 2cm 1cm},clip,width=\linewidth]{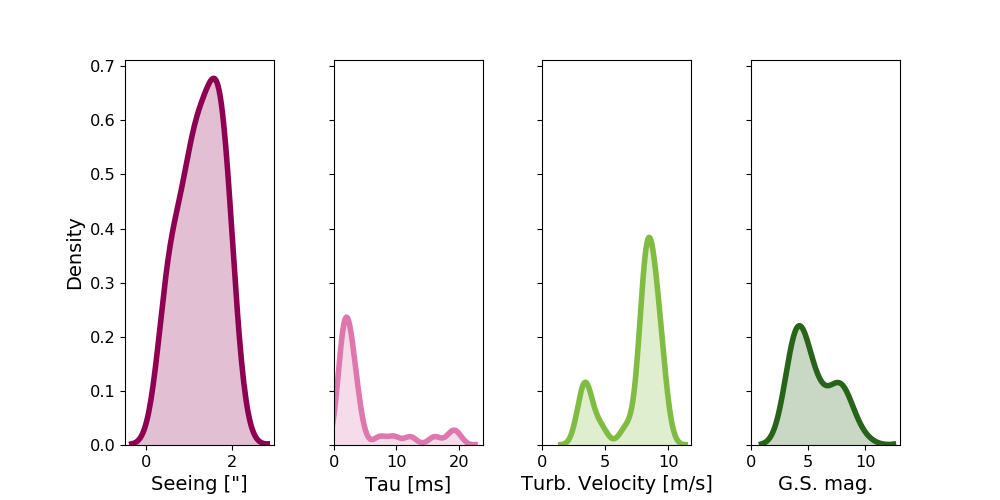}
\caption{Kernel density functions for the seeing, coherence time, turbulence velocity, and guide star magnitude (r band) showing the conditions under which the VLT/SPHERE telemetry was taken. The first three panels are measurements closest to the time of the observation output by the Adaptive Optics Facility (AOF; accessed via ESO Paranal query form). The corresponding targets were found at the VLT/SPHERE ESO archive and their r-band magnitudes were found in the VizieR catalog. The kernel density functions are a nonparametric estimation of the probability functions. }
\label{fig:conditions}
\end{figure}
\begin{figure}
\centering
\includegraphics[trim={0.5cm 0cm 0.5cm 0.5cm},clip,width=0.8\linewidth]{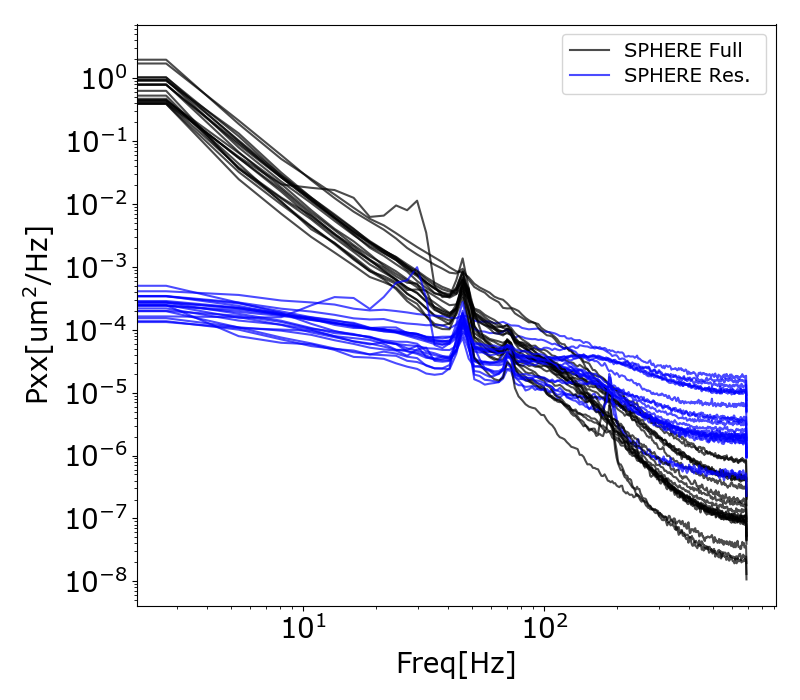}
\caption{Power spectral densities,  {estimated using the Welch method}, for all the data sets; both for the full VLT/SPHERE {estimated} pseudo open-loop phases and for the reconstructed VLT/SPHERE residual phases. }
\label{fig:all_psd}
\end{figure}

The SAXO system has the option to save the full (or partial) XAO telemetry, including HODM positions, SHWFS slopes, SHWFS intensities,  {interaction} matrix, at the discretion of the instrument user.  In this paper, we make use of 27 SAXO data sets taken between 2017 and 2019. By limiting ourselves to these years we also have estimations of the atmospheric conditions (seeing, coherence time, and turbulence  {velocity}) from the MASS-DIMM instrument located approximately 100 meters away from UT4, which houses the VLT/SPHERE instrument. The conditions under which our data were acquired are summarised in Fig.~\ref{fig:conditions}, where we plot the kernel density functions estimated from the data. We note that the data set is biased toward shorter coherence times (tau), where the median coherence time are 2.5 milliseconds; from \citet{Milli_2017} and \citet{Cantalloube_2018} we expect the WDH to be dominant (and thereby the servo-lag error) when the coherence time drops below 5 milliseconds. For completeness, the data set has  {a couple of data sets with} longer coherence times. The turbulence velocity is the velocity of the  {characteristic} turbulent layer as determined by the MASS-DIMM instrument  {and is associated with a characteristic altitude determine from the atmospheric profile. Therefore, it does not necessarily indicate the speed of the jet stream layer but provides a tracer atmosphere velocity.}  For most of the data we have  {bright} guide star magnitudes with a mean of 5 magnitudes in the r band (the wavefront sensor bandwidth), resulting in  {high} signal-to-noise ratios (S/N) for the SHWFS in all cases. The seeing has a Gaussian-like distribution with a mean of 1.3 arcseconds. A full summary of the entire data set (including time of observation) is provided in Tab.~\ref{tab:data} in the Appendix. Each of the 27 data sets differ in length and range from 10 s to 60~s, allowing us to probe different conditions while still having the opportunity to observe the behaviour of turbulence on timescales of a minute. 

To study the influence of prediction on our data sets, we {estimated} the pseudo-open loop phases, thereby applying prediction to the zonal two-dimensional grid of phase values and not the modes. We performed the open-loop {estimation} using the HODM commands only because SAXO saves the full HODM voltages, not just the updates {(see Appendix~\ref{ap:2})}. We converted to phase using the standard SAXO HODM modes. As a result, we neglected the spatial frequencies beyond the spatial bandwidth of the HODM and therefore underestimated the open-loop phase at higher spatial frequencies. We took this approach after first considering the more traditional method of using the wavefront sensor measurements and unravelling the pseudo open-loop phase from the SHWFS measurements and knowledge of the controller state.  Without full knowledge of the modal gains and controller at each time step, as in our case, this method provides an inaccurate {estimation}.

In Fig.~\ref{fig:all_psd}, we plot the resulting PSDs for the final pseudo open-loop phase from the VLT/SPHERE telemetry (SPHERE full) and the closed-loop phase of VLT/SPHERE residuals and identify some key features. There are peaks around 40 hertz and 60 hertz in both the open and closed-loop PSDs. Performing a modal analysis we find that the peaks appear for Zernike modes 7-11 \citep[Noll indexing;][]{Noll_1976} with varying amplitudes. The second feature is the increase in power at high temporal frequencies for the closed-loop PSD  (i.e., the so-called waterbed effect, a result of Bode's sensitivity integral).
\fi
\section{Methodology} 
\label{sec:methdology}
 Although we are ultimately interested in the improvement in contrast using prediction, we limit ourselves in this work to studying the minimisation of the servo-lag error. From~\citet{Guyon_2005},  { ~\cite{Kasper_2012}, }and ~\citet{Cantalloube_2018}, we see that minimising the lag results in an improvement in contrast  {but ultimately also} depends on the coronagraph of choice  {and how it interacts with the residual phase at small angular separations. We also do not have focal plane images taken at the same time as the data sets, making a clear claim of improvement unreliable}. The metric we adopt is the temporally and spatially averaged wavefront phase variance. 
\subsection{LMMSE prediction} 
\label{sec:LMMSE}
\iftrue

For our predictor we chose a data-driven method called the LMMSE predictor. This approach provides a flexible framework that allows us to implement the predictor in three different ways: batch, recursive, and with an exponential forgetting factor \citep{vanKooten_2019}. 

A single point $i$ of a phase screen at time $t$ is given by $y_i (t)$, while $\vec{u(t)}$ is a $P^2 \times$1 column vector containing a collection of $P^2$ phase values on a discrete spatial grid at time $t$. We assume that the future value of a given phase point, $\hat{y}_i$ at the discrete time index $t+d$, is a linear combination of the most recent phase values at time $t$. The predictor coefficients are denoted as $\vec{a}_i$. The cost function of our predictor, with $ < > _t$ as the time average operator, is then
\begin{equation}
min_{\vec{a}_{i}} <||{y}_i(t+d)-\vec{a}_i ^{T} \vec{w}(t)|| ^2 >_t ,
\label{eq:mmse}
\end{equation}
 
 where $\vec{w(t)}$ includes a set of $Q$ most recent measurements,

\begin{equation}
\vec{w(t)}=
 \begin{pmatrix}
  \vec{u}(t)^{T}  & \vec{u}(t-1)^{T} & \vec{u}(t-2 )^{T} & ...& \vec{u}(t-Q)^{T} 
 \end{pmatrix}^{T} .
  \label{eq:w}
\end{equation}
 
We allowed for both spatial and temporal regressors gathered into $\vec{w(t)}$ -- a $P^2 Q\times$1 vector. We denoted predictors of various orders by indicating the spatial order $P$ (spatially limiting ourselves to a box of order $P$, symmetric around the phase point of interest, resulting in $P^2$ spatial regressors) followed by the temporal order $Q$; for example, a `s5t2' predictor has $P=5$ and $Q=2$.

Solving Eq.~\ref{eq:mmse} for our zero-mean stochastic process, the solution can be written in terms of the inverse of the auto-covariance matrix and cross-covariance vector~\citep{Haykin_2002}
\begin{equation}
\vec{a}_i=\mathbf{C}_{\vec{w}\vec{w}}^{+}\vec{c}_{\vec{w}{y}_i}
\label{eq:batch}
,\end{equation}
where $+$ denotes a pseudo inverse; $\mathbf{C}_{\vec{w}\vec{w}}$ is the auto-covariance matrix of $\vec{w}$, the vector containing the regressors; and ${c}_{\vec{w}{y}_i}$ is the vector containing the cross-covariance between the true phase value, ${y}_i$ and $\vec{w}$. 

We can estimate the covariances in Eq.~\ref{eq:batch} directly from a training set, forming a fixed batch solution. Alternatively, we can form a recursive solution making use of the Sherman-Morrison formula (a special case of the Woodbury matrix inversion lemma).  {In Eq.~\ref{eq:update_cc} through to Eq.~\ref{eq:cov2} } we inserted the exponential forgetting factor in the update in the following equations, thereby forming our final LMMSE implementation. The recursive form is given in Eqs.~\ref{eq:update_cc} and~\ref{eq:cov1} with $\lambda=1,$ such that all previous data is weighted equally, as follows:   

\begin{equation}
\vec{c}_{\vec{w}{y}_i}(t-d)=\lambda \vec{c}_{\vec{w}{y}_i}(t-d-1)+\vec{w}(t-d){y}_i(t-d)
\label{eq:update_cc}
\end{equation}

\begin{equation}
\mathbf{C}_{\vec{w}\vec{w}}^{+}(t-d)=\lambda ^{-1}\mathbf{C}_{\vec{w}\vec{w}}^{+}(t-d-1)-
\vec{k}(t-d)
\label{eq:cov1}
\end{equation}

with
\begin{equation}
\vec{k}(t-d)=\frac{\lambda ^{-2}\mathbf{C}_{\vec{w}\vec{w}}^{+}(t-d-1) \vec{w}(t-d)\vec{w}^T(t-d)\mathbf{C}_{\vec{w}\vec{w}}^{+}(t-d-1)}{1+\lambda ^{-1}\vec{w}^T(t-d)\mathbf{C}_{\vec{w}\vec{w}}^{+}(t-d-1)\vec{w}(t-d)}
\label{eq:cov2}
.\end{equation}
By updating Eqs.~\ref{eq:update_cc} and~\ref{eq:cov1}, the coefficients can be found for each time step using Eq.~\ref{eq:batch}.
 The recursive solution goes on-line immediately with the initial auto-covariance being set to diagonal matrices with large values (as done with recursive least-squares methods) and the cross-covariance vector set to ones. By adjusting the forgetting factor, we can weigh old data by less compared to the most recent measurements, therefore allowing the tracking of slowly varying signals. 
\fi

\subsection{Comparison with EOF}
\iftrue

Methods such as LMMSE and EOF (see \citet{Guyon_2017}) and similar techniques (see \cite{Jensen-Clem_2019}) all minimise the same cost function, but the evaluation of Eq.~\ref{eq:batch} is different in each case; we note that the cost
function is slightly different when including exponential forgetting factor. In EOF, the solution is estimated with the inverse of the auto-covariance determined using a singular-value-decomposition that is re-estimated on minute timescales. The amount of data used to estimate the prediction filter, the numerical robustness, the noise properties of the system, and the atmospheric turbulence above the telescope all contribute to the performance of these algorithms and the final computational load. Therefore one implementation might be more suited for specific conditions than the other, but the three methods can, in ideal conditions, result in the same performance.
\fi
\begin{sidewaysfigure*}
\centering
\includegraphics[trim={1cm 0cm 0cm 3cm},clip,width=0.8\textwidth]{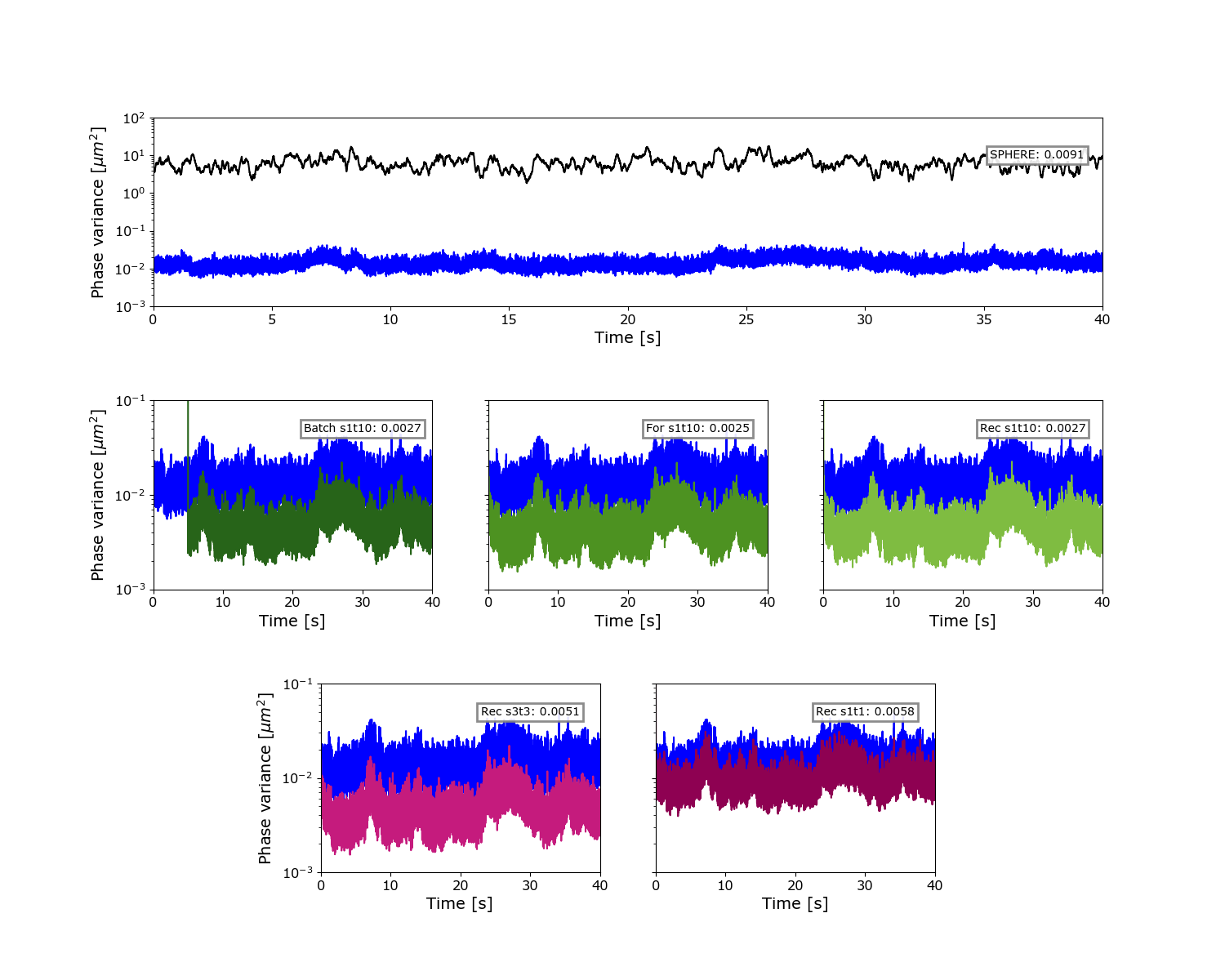}
\caption{In the top panel, we plot a time series showing the {estimated} pseudo open-loop in black, compared with the VLT/SPHERE residual in blue for a random SAXO telemetry data set. The other plots (all with the same y-scale) show various predictor residual phase variances (various green and purple lines) compared to the VLT/SPHERE residuals for the same data. The averaged phase variance, in $\mu m^2$, for the last 5 s is indicated in the top right corner of each plot. The forgetting s1t10 (abbrev. for) does not perform significantly better than the recursive s1t10 (abbrev. res). Secondly, the s3t3 does not perform better than the s1t10.}
\label{fig:timeseries}
\end{sidewaysfigure*}

\subsection{Applying prediction to SAXO telemetry}
\label{sec:applying}
\iftrue

\begin{sidewaysfigure*}
\centering
\includegraphics[trim={1cm 0cm 2cm 0cm},clip,width=0.9\textwidth]{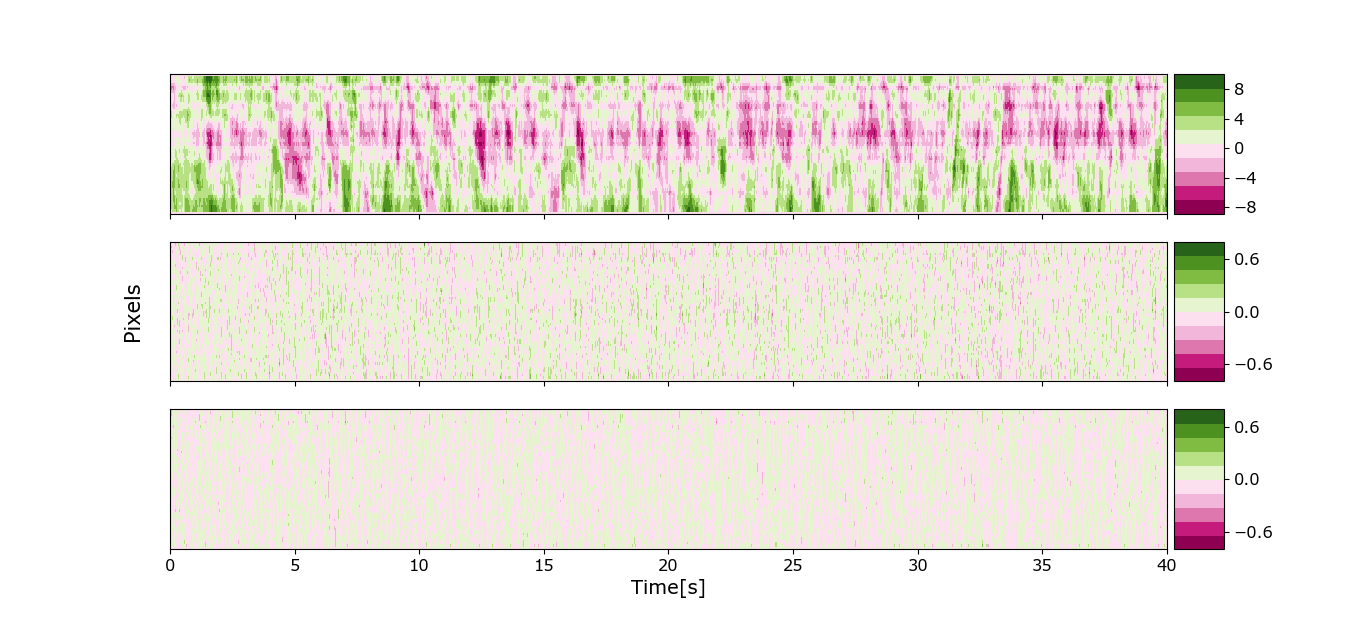}
\caption{ Vertical slice across the telescope aperture (y-axis) showing the wavefront phase in $\mu$m (indicated by the colour-bars), plotted as a function of time for the pseudo open-loop phase (top), the VLT/SPHERE residual phase (middle), and s1t10 predictor residual phase (bottom) for the same night as Fig.~\ref{fig:timeseries}. Comparing the bottom two panels (colour map is the same in both), we can see the prediction residuals have a flatter and more uniform appearance compared to the real VLT/SPHERE residuals. }
\label{fig:radial}
\end{sidewaysfigure*}

From the {estimated} open-loop phases (which results in a 240-by-240 phase screen using the HODM modes) we bin the data to 60-by-60 phase screens for computational memory purposes. We then performed prediction on the estimated open-loop phases assuming a 2-frame delay. For each phase point, we estimated a unique set of prediction coefficients using the equations as outlined in Sec.~\ref{sec:LMMSE}. Our aim is to focus on the prediction capabilities, ignoring the control aspect by assuming a perfect system -- no wavefront sensor noise and a HODM that can perfectly correct all spatial frequencies predicted -- and only including the delay.  We note that this results in no fitting error, no spatial bandwidth limitations, and no temporal bandwidth limitations on the achievable performance.

We ran batch, recursive, and forgetting (with $\lambda = 0.998$ as this value gives the best performance assuming $\lambda \neq 1$) LMMSE predictors for each prediction order. We then subtracted the predicted phase from the pseudo open-loop phase 2 frames later resulting in the residual predictor phase. From the phases we calculated the averaged phase variance by taking the final 5 s of the data set, and therefore, all the different predictors have converged; we calculated the spatio-temporally averaged phase variance. 
We started with the performance of a s1t1 predictor. The s1t1 is a zero-order predictor because it only makes use of the most recent measurement for that given phase point, making it analogous to an optimised integrator (with a gain close to unity) in our simulations; we refer to the s1t1 as the ideal VLT/SPHERE performance. 

We  performed simulations testing a variety of predictors with different spatial and temporal orders including s1t3, s1t10, s3t1, and s3t3. We looked at how the three different implementations of the LMMSE for each prediction order behave (see Fig~\ref{fig:timeseries}). In Sec.~\ref{sec:results} we present the results for the recursive s1t10 predictor, which performs the best out of all the different orders. 
\fi


\section{Results}
\label{sec:results}

\begin{figure*}
\centering
\includegraphics[trim={1cm 0cm 1cm 0cm},clip,width=0.8\linewidth]{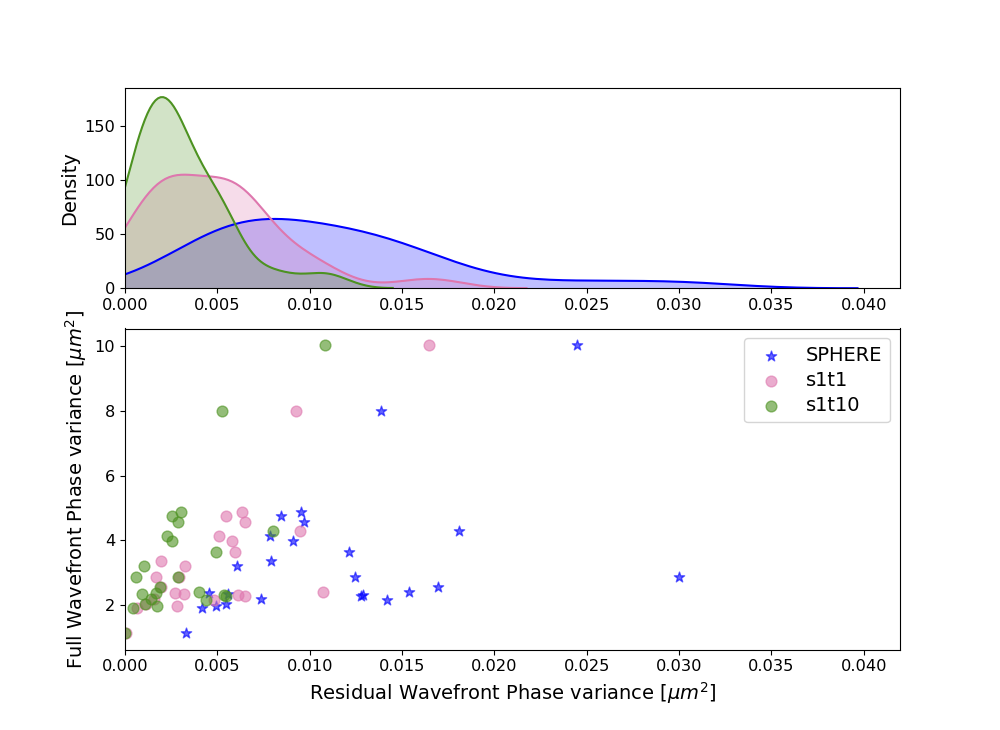}
\caption{Top: Kernel density function estimation for the averaged phase variances plotted in the bottom plot. The change in spread of the values is shown. Bottom: Pseudo open-loop averaged wavefront phase variance compared to residual averaged wavefront phase variance for VLT/SPHERE, a batch s1t1 (i.e., idealised integrator for VLT/SPHERE), and a recursive s1t10 predictor. The points move from right to left, indicating that a s1t1 does better than VLT/SPHRE and a high order predictor does even better than the s1t1. }
\label{fig:fullvsres}
\end{figure*} 
We find that prediction provides a reduction in averaged phase variance when compared to the VLT/SPHERE SAXO residuals. An example of how prediction behaves spatially and temporally for a slice across the telescope aperture is shown in the bottom panel of Fig.~\ref{fig:radial}. We see a reduction in the phase compared to the VLT/SPHERE residuals and see a more uniform solution in time and space. In Fig.~\ref{fig:fullvsres} we summarise the results of running prediction on all of our data sets. As mentioned above, the averaged phase variance is found by using the last 5 s of the data. The data sets all vary in length.  {When we see a reduction in the residual phase variance, we would expect an improvement of performance for the XAO system} for all cases independent of the guide star magnitude, the seeing, and the coherence time. We  observe a reduction in the spread of residual phase variance, with prediction providing a more uniform performance for various conditions; see the kernel density function plots in the top panel of Fig.~\ref{fig:fullvsres}. 

\begin{figure*}
\centering
\includegraphics[trim={1cm 0cm 1cm 0.1cm},clip,width=0.8\linewidth]{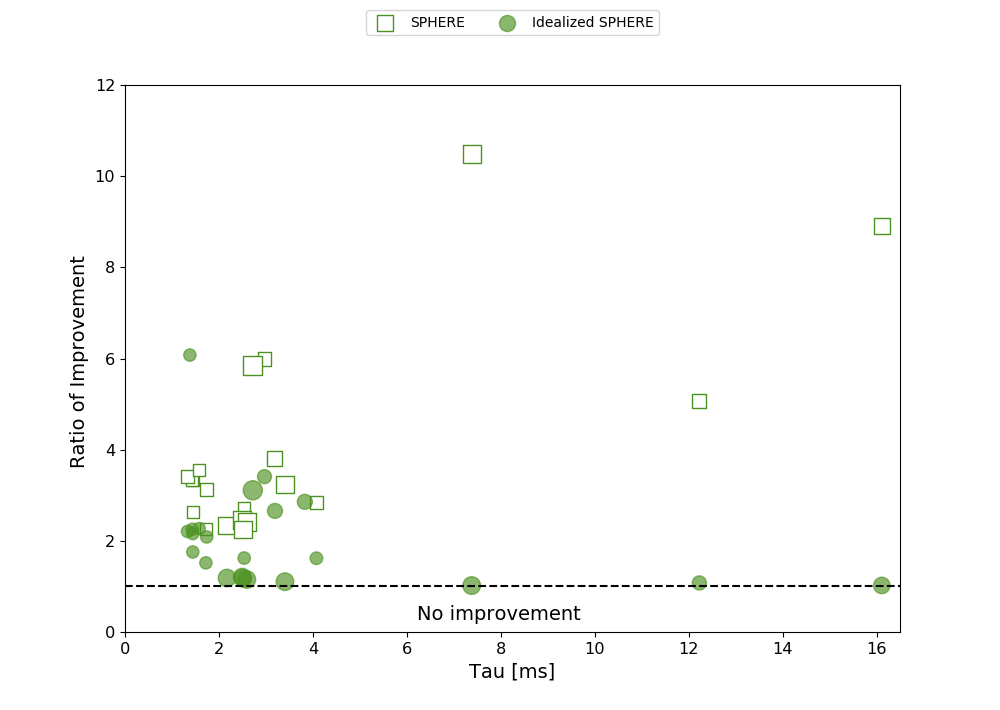}
\caption{Ratio of improvement, found by taking the ratio of an idealised integrator on VLT/SPHERE to a recursive spatial-temporal predictor (s1t10) phase variance as calculated from the last 5 s of data, as a function of coherence time. The size of the markers indicates the AOF seeing conditions at the time of observation. The average ratio of improvement is 5.1 when comparing the prediction to the real VLT/SPHERE residuals. When looking at the idealised VLT/SPHERE, we find an average ratio of improvement of 2.0 in wavefront variance reduction.}
\label{fig:ratio}
\end{figure*}

In Fig.~\ref{fig:ratio} we plot the ratio of the recursive s1t10 predictor residual phase variance to the VLT/SPHERE residual phase variance defining this as the ``ratio of improvement'' as a function of coherence time. In the same figure, we add the ratio of improvement for the same recursive s1t10 predictor and an idealised VLT/SPHERE integrator (batch s1t1) against coherence time. We calculate the average ratio of improvement to be 5.1 and 2.0, respectively. Fig.~\ref{fig:ratio} shows the relative seeing conditions indicated by the size of the marker; smaller marker sizes indicate better seeing conditions. 

\begin{table*}
\centering
 \begin{tabular}[c]{||c|c|c|c|c|c||}
 \hline
 Time of data & {Pseudo} open-loop [$\mu m^2$]\ & VLT/SPHERE [$\mu m^2$] & s1t3 [$\mu m^2$] & s3t3 [$\mu m^2$] &s3t1 [$\mu m^2$] \\ [0.5ex] 
 \hline\hline
 2017-07-19T22:54:55.000 &  1.963& 0.005        &0.002
& 0.002 & 0.002
 \\ 
 \hline
 2017-07-19T23:16:39.000 &   2.856&     0.012&  0.001&
 0.001 & 0.002 \\
 \hline
 2017-07-19T23:22:02.000 &2.400&        0.015&  0.004&
0.004 & 0.011\\
 \hline
 2017-07-19T23:50:33.000 & 2.148        &0.014& 0.004&  0.004 & 0.005
 \\ 
 \hline
\end{tabular}
\caption{Averaged phase variance for the  pseudo open-loop, VLT/SPHERE residuals, s1t3 residuals, s3t3 residuals, and the s3t1 residuals. Comparing the last three columns, there is no gain by including spatial regressors to the prediction algorithm and the s1t3 does better than the s3t3.}
\label{tab:variance}
\end{table*}

We evaluate several predictors, varying both spatial and temporal orders. We find that there is no gain in performance by adding spatial regressors and find that temporal regressors perform equally as well. These results are summarised in Tab.~\ref{tab:variance}. 

We see minimal evidence of nonstationary behaviour of optical turbulence. First, by looking at the coherence times in Tab.~\ref{tab:data}, we do not see significant change in coherence time for data taken on the same night -- showing that on 100 s time scales the statistics of optical turbulence do not vary significantly.  Second, on shorter timescales, we do not see evidence of nonstationary turbulence. In Fig.~\ref{fig:timeseries}, we plot the batch, recursive, and the forgetting LMMSE for a s1t10 predictor. From the average residual phase variances, we see that the batch and recursive perform the same over the full 40 s period. We do see a slight improvement for the forgetting LMMSE, implying a slight time-variant behaviour of the pseudo open-loop phase but nothing significant. 
    
    
 

\section{Discussion}
\label{sec:discuss}
\subsection{Comparison of the prediction residuals to the VLT/SPHERE residuals}
\label{sec:comparison}
We should note the difficulties in performing a direct comparison between the predictor residuals to the real VLT/SPHERE residuals. There are a few challenges, with the first being a difference in delay. In our {estimation of the open-loop phase} we choose to round the frame delay to a whole frame; therefore our predictor sees a delay of 2 frames (or 1.45 ms) while the real system delay, and thereby encoded in the real VLT/SPHERE residuals, is 2.2 frames (1.59 ms). We therefore assume that the VLT/SPHERE residuals immediately have a larger phase variance compared to the case where the true delay is 2 frames.  An alternative option to rounding the delay frames is to interpolate, which is a step that would have also introduced an error. We make use of the HODM commands which are saved as the total voltage on the HODM, not the update to the HODM. The SHWFS, however, is used to determine the real VLT/SPHERE residuals and therefore see higher order spatial frequencies, potentially increasing the residuals if this was not the case. Perhaps the most substantial contribution to the final performance of VLT/SPHERE results from the fact that the VLT/SPHERE performance will be limited by operational parameters. The controller will need to be stable and robust on-sky, potentially resulting in a loss of performance when compared to our idealised situation. Therefore, in Figs. ~\ref{fig:fullvsres} and ~\ref{fig:ratio} we plot the idealised VLT/SPHERE (batch s1t1) as well as the VLT/SPHERE residuals. The true gain in prediction will be between the idealised VLT/SPHERE and real VLT/SPHERE residuals and the ratio of improvement will fall between 5.1 and 2.0. 

\subsection{Performance under different conditions} \label{sec:conditions}

\begin{figure*}
\centering
\includegraphics[trim={1cm 0cm 1cm 0.1cm},clip,width=0.9\linewidth]{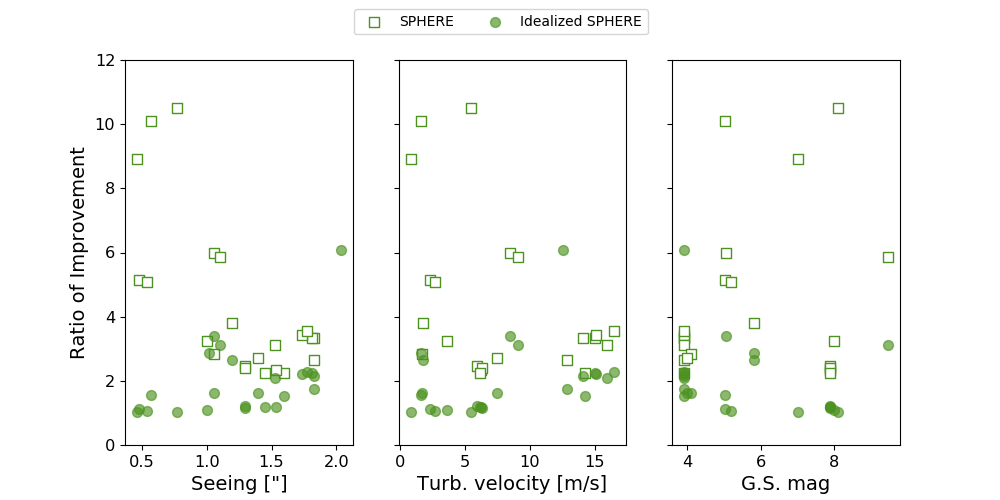}
\caption{Ratio of improvement compared to seeing (left), turbulence velocity (middle), and guide star magnitude in r band (right) during the time of observation.}
\label{fig:ratio_all}
\end{figure*}

Our data set and analysis is unique, showing that with prediction, there will be an improvement under almost all conditions even for long coherence times, and no loss in performance with prediction is observed. However, the data set does not show any clear correlations between predictor performance and observational conditions (see Figs.~\ref{fig:ratio} and~\ref{fig:ratio_all}). We briefly discuss the behaviour for various observing parameters including coherence times, turbulence velocity, seeing, and finally guide star magnitudes. 

Looking closely at Fig~\ref{fig:ratio} and the behaviour for various coherence times we do not find any correlation between the coherence time and performance for the true VLT/SPHERE residuals. However, when looking at the idealised VLT/SPHERE behaviour we see, at smaller coherence times, an exponential-like gain in the ratio of improvement in which the asymptote is the Nyquist sampling frequency ($2/WFS_f=2/1380=1.45 ms$). This behaviour for the idealised case is as expected, where we have a larger improvements for shorter coherence times.  We then look at the relation between the ratio of improvement and turbulence velocity (middle plot of Fig.~\ref{fig:ratio_all}). We expect a similar behaviour to that of the coherence time as they are related. We note that we also see no correlation for the ground layer wind speeds measured by the nearby meteorological tower. Studying the relation between the seeing and the ratio of improvement we notice an asymptotic behaviour where the ratio improves for better seeing conditions. We do not see any dependence between performance and S/N (or guide star magnitudes; left plot of Fig.~\ref{fig:ratio_all}).  For the VLT/SPHERE residuals, the lack of correlation between performance and S/N is as expected from laboratory and on-sky validation of SAXO by ~\cite{Fusco_2016} at these guide star magnitudes. We also note that for long coherence times VLT/SPHERE is often looking at fainter targets since the conditions are ideal. This is the case for our data as well (see Tab.~\ref{tab:data}). In summary, we see a relation between the true VLT/SPHERE residuals and the seeing, but for the idealised VLT/SPHERE case we see a relation between the improvement and the coherence time, as expected. The lack of correlation between ratio of improvement and the other observing parameters could be a result of the different locations of VLT/SPHERE and DIMM-MASS (which measures the observing parameters) and that the values determined over minute averages do not reflect the exact values at the time of observation. Alternatively, the behaviour of the SAXO controller is limited by internal system requirements (such as vibration rejection) and not the observing conditions, resulting in a lack of correlation between observing conditions and gain in performance. 

Studying Fig.~\ref{fig:ratio}, we can see that the ratio of improvement when using idealised VLT/SPHERE as a benchmark behaves very differently than the true VLT/SPHERE case. For a direct comparison to VLT/SPHERE, we do not see any correlations between the ratio of improvement and the coherence time. However, for the idealised case, we see that at longer coherence times we have no improvement since the SAXO XAO system can already perform well.

\subsection{Time-invariant turbulence statistics}\label{sec:time_invariant}

Observing the behaviour of the three (batch, recursive, and exponential forgetting) implementations of the LMMSE, we can comment on the stationarity of the turbulence. The LMMSE finds the optimal prediction coefficients determined from the training set. The batch is trained on the first 5 s and the recursive trains continuously. We note the residual phase variances for the last 5 s for each case; we do not see a significant difference between the batch and recursive solutions in any case, indicating the statistics of the turbulence has not changed over the measurement period. Studying the recursive solution, we look at the behaviour of the prediction coefficients in time across the aperture. We do not see any notable changes over the entire period once the solution has converged. We do see more fluctuations in the prediction coefficients for phase points located on the edges with a predictor using spatial information such as the s3t3. Conversely, the exponential forgetting LMMSE does show a slight improvement, however, this could be due to noise in the system that the LMMSE predictor can and does remove.  

\subsection{Implications of results}
\label{sec:implications}

For all conditions we see an increase of performance with prediction when compared to the VLT/SPHERE residuals, under an idealised assumption there are a few cases where the ratio of improvement is one, indicating no gain but also, notably, no loss in performance. 

A more notable result is from the kernel density functions plotted in the top panel of Fig.~\ref{fig:fullvsres}. The spread of the averaged phase variance is less for the s1t10 predictor, indicating a more uniform performance in phase variance reduction for different observing conditions. From an observational point of view, having a more stable correction under different conditions is desirable, especially for the cases on surveys in which observers might be targeting similar objects and can perform reference star differential imaging from a library.   

The ratio of improvement we find is 5.1, but this is probably an overestimate of the improvement we could achieve on-sky. In previous prediction work,~\citet{Guyon_2017} show an improvement of 7 in root-mean-square (rms) residual wavefront error while offline telemetry tests by~\citet{Jensen-Clem_2019} show a factor of 2.5 rms wavefront error; we note that the errors in both of these works refer to systems located on top of Mauna Kea. Although an exact direct comparison using these values is impossible, we note that we show a more modest predictive improvement compared to these studies. 

 When studying the prediction order we do not see a large gain from including spatial information.  This is due to large sub-aperture size and the high rate of temporal sampling, which results in the turbulence only moving across a sub-aperture after many frames; for example, assuming 10 m/s and 0.2 m sub-aperture size, it takes approximately 28 frames before a cell of turbulence moves to the next sub-aperture.  The turbulence is still dynamic but we sense the average variations that fall within the wavefront sensor sub-aperture. We have much finer temporal sampling. We expect a temporal-only predictor to be best suited for HCI, and by removing the spatial information, we are no longer sensitive to wind direction (spatial solution requires a symmetric choice of regressors) which therefore reduces the computational size of the prediction problem.   

We do not see evidence of time-invariant turbulence, meaning that our predictor does not need to be able to track changes in turbulence behaviour on timescales less than 1-2 minutes of observations. We see a slight increase in performance from the exponential forgetting factor LMMSE solution but the difference is very slight and not substantial enough to suggest this as the best choice. From a computational point of view, the batch LMMSE is the best option and resetting it every 1 to 2 minutes (or as needed based on longer telemetry data) using 5 s of training data would be the best implementation. 

    
    
 
 

\section{Conclusions and future work}
We find a reduction in the phase variance in comparison to the VLT/SPHERE residuals and determine the ratio of improvement to be 5.1 for SAXO telemetry data. When prediction is compared to an idealised VLT/SPHERE system, we find an improvement ratio of 2.0. In all cases, no matter what the observing conditions, prediction performs well with no loss in performance. Most importantly, we note that under all the 27 various observing conditions studied, we see a reliable and overall more consistent improvement of the system performance. The data set, in combination with our predictors, reveals that the optical turbulence as seen by the telescope is time-invariant and that the temporal regressors have a larger impact on the performance of the predictor than spatial regressors. We recommend a batch (updating every few minutes as necessary) temporal-only predictor for the VLT/SPHERE to reduce the servo-lag error. In future work we will seek to investigate the effects of prediction on contrast for different types of coronagraphs and to determine whether predictive control can be used to directly optimise the raw contrast.  
\label{sec:conclusion}



\begin{acknowledgements}
      The authors would like to thank Markus Kasper and Julien Milli for providing us with the VLT/SPHERE SAXO data. The authors would also like to thank Leiden University, NOVA, METIS consortium, and TNO for funding this research.
\end{acknowledgements}

%
   \bibliographystyle{aa} 
   \bibliography{report} 
%
\onecolumn
\appendix
\section{Overview of SAXO data}
In this Appendix we provide a full summary of our data set. The XAO telemetry data is stored on the SAXO server and a log of when AO data was taken can be found on the ESO science archive under the VLT/SPHERE instrument. The 2019 data was kindly provided to us by Markus Kasper while the other data sets were accessed by Julien Milli. 
\begin{table*}[!h]
\centering
\begin{adjustbox}{width=0.85\hsize}
\begin{tabular}{||c|c|c|c|c||}
\hline
    \bfseries Date & \bfseries Seeing["] & \bfseries Tau[ms]  & \bfseries Turbulence Velocity [m/s] &\bfseries Guide star Magnitude r-band
    \begin{filecontents*}{SPHERE_data_used.csv}
Date,Seeing [''],Tau[ms],Mass-Dimm turbulence velocity[m/s],METEO 10 m wind [m/s],METEO 30 m wind [m/s],Target (ESO Archive),R-band mag (simbad,Mean variance full turbulence,Mean variance rec turbulence,Mean variance s3t3 rec,Mean variance s1t10 rec,Mean variance s3t3 batch,Mean variance s1t1 rec
2017-07-17T23:15:30.000,1.708,1.539,8.340,14.400,14.880,-,3.910,0.000,0.000,0.0029806492555413700,0.0028270268673055700,0,
2017-07-17T19:07:58.000,2.037,1.381,-,12.030,12.530,PSF calibration,3.910,1.1321361650339700,0.0032834510008532000,0.00021262034083253000,5.7022542784927E-06,0.034389550347419,3.46497297572297E-05
2017-07-17T23:00:24.000,1.829,1.442,8.450,15.200,15.150,HD 110458,3.910,0.000,0.000,0.000,0.00000,0.004,0.007
2017-07-17T23:02:51.000,1.827,1.443,8.450,12.780,12.900,-,3.910,7.981848459072760,0.013879316153939900,0.0053101706643945400,0.005266932015125810,0.005764805048451,0.009238356445922900
2017-07-17T23:04:41.000,1.825,1.444,8.450,13.880,14.100,-,3.910,4.747400717361380,0.008443004213262810,0.0024943748951342800,0.002532820694638510,0.002830852464367,0.005465141603769690
2017-07-17T23:10:31.000,1.812,1.434,8.450,14.950,15.050,-,3.910,4.5753524411347000,0.009664909148582010,0.0028529063963572700,0.002893385326357500,0.003296409467313,0.006515432951170220
2017-07-17T23:11:54.000,1.739,1.330,9.210,14.100,15.130,-,3.910,4.14557206561174,0.007838240622927370,0.0022693603704559300,0.0022965026595691800,0.002470523717541,0.005063955321791900
2017-07-17T23:29:56.000,1.776,1.579,9.280,15.800,16.520,-,3.910,3.9765489041049000,0.009093766852504200,0.00252572383715147,0.002560322270061840,0.002809005819498,0.005808037547956620
2017-07-17T23:31:11.000,1.599,1.722,8.300,13.530,14.280,-,3.910,10.031656667749900,0.02448637622360470,0.010862099997241000,0.010852592273411700,0.012826868206491,0.0164474607475725
2017-07-17T23:33:29.000,1.524,1.738,7.970,15.300,15.950,-,3.910,4.883391275821880,0.009546512904766830,0.0030156625093809100,0.0030512538548626000,0.00327804779105,0.0063578749769976200
2017-07-19T22:54:55.000,1.053,4.075,7.990,1.350,1.730,HD 129116,4.110,1.9624998805781800,0.004943139646326480,0.0017141014183910900,0.0017434141163331800,0.002035176597865,0.0028188408509284200
2017-07-19T23:16:39.000,1.014,3.829,9.120,0.750,1.600,HD 118646,5.819,2.855977534504140,0.012444610608562600,0.0005960008747611640,0.0005794813508222300,0.000909153716561,0.001654697030481550
2017-07-19T23:22:02.000,1.197,3.193,8.760,0.930,1.770,-,5.819,2.399834382818660,0.01535262294661980,0.004081493247515250,0.004032754014961500,0.00590660746225,0.010709256914158400
2017-07-19T23:50:33.000,1.001,3.406,9.700,3.830,3.630,HD 132347A,8.010,2.147839662599720,0.014159887967670700,0.004305819470059160,0.004374390213560560,0.004601067216778,0.004821383743767940
2018-04-04T03:06:15.000,1.059,2.970,0.000,8.150,8.480,HIP 21547,5.060,2.3295914221979800,0.0055665678276071500,0.000868624546427019,0.0009297832331594750,0.001374430960232,0.0031689232650442300
2018-04-04T03:12:50.000,1.099,2.721,0.000,8.820,9.070,TYC 7617 0549 1,9.480,3.2075118677608600,0.006082971555933840,0.0009975772826377130,0.0010385827595652200,0.001550843128353,0.0032287283136842500
2016-05-21T08:23:42.000,1.396,2.538,6.65,6.95,7.47,HIP_95347,4.00,2.377161214385540,0.004525363658596300,0.0016979986553178300,0.0016616766640963400,0.001773524713549,0.0026937028844889200
2016-05-21T09:29:07.000,1.533,2.171,8.16,6.18,6.25,HIP_102409,7.890,2.2676366453200900,0.012752421309030300,0.005469403467963280,0.005472929527397130,0.005862791300754,0.006476660499099500
2016-05-21T09:53:34.000,1.298,2.495,9.31,5.85,5.93,HIP_102409,7.890,3.6362386152474200,0.012116361192908400,0.004919262239344880,0.0049388998593951100,0.005447355804406,0.00594631318239201
2016-05-21T09:55:48.000,1.292,2.596,9.03,6.28,6.33,HIP_102409,7.890,2.3071639268125800,0.012881648020362400,0.005305306148130390,0.005340662871957200,0.005432011701382,0.006136762053552030
2016-05-21T09:58:02.000,1.452,2.510,8.31,6.58,6.20,HIP_102409,7.890,4.294696793759530,0.018081321523302300,0.008036689831465700,0.008037026235410550,0.009001958165081,0.009459295950207820
2019-01-26T01:49:48.00,0.770,7.381,3.39,5.63,5.50,HIP_20130,8.110,2.8625183031112000,0.030015095422388800,0.0029359024326736100,0.002860920018697,0,0.002909020040961
2019-01-26T02:08:25.00,0.710,9.596,3.39,4.15,3.92,HD 19467,6.560,3.349309210344830,0.007899907671529080,0.001839997206102810,0,0,0.001952261523916
2019-01-26T03:34:39.00,0.460,16.113,4.65,0.63,0.82,HIP014898,7.017,2.5615343407281400,0.016938554307024500,0.001968103233106510,0.001901405386992,0,0.001939222434773
2019-01-26T03:48:01.00,0.570,19.643,3.1,1.5,1.65,HIP50847,5.020,1.8952848814185900,0.004164301351509100,0.0005229924260328780,0.000412642131913,0,0.000647210297394
2019-01-26T04:59:53.00,0.480,18.774,3.67,2.28,2.35,HIP50847,5.020,2.1923817094380600,0.007344874686360350,0.001511551872208180,0.001429166996293,0,0.001591690532763
2019-01-26T06:39:57.00,0.540,12.228,7.81,2.15,2.73,HIP52742,5.180,2.039421237331780,0.005472769595685240,0.0011734566301666400,0.00107976222409,0,0.001160280450313
\end{filecontents*}
\csvreader[%
     head to column names,after head=\\ \hline,late after line=\\\hline]{SPHERE_data_used.csv}{}
    {\csvcoli&\csvcolii &\csvcoliii &\csvcoliv &\csvcolviii}

    \end{tabular}
    \end{adjustbox}
\caption{Summary of the VLT/SPHERE data used in this work as well as the AOF atmospheric conditions as recorded closest to the measurement time. The target r-band magnitude is provided as to provide a relative estimation of the signal-to-noise ratio for the wavefront sensor. The length of the telemetry data ranges from 10 to 60 seconds.}
\label{tab:data}
\end{table*}
\section{{Estimation} of open-loop phase}
\label{ap:2}
 {In this Appendix, we explain how the SAXO VLT/SPHERE telemetry data is used to {estimate} the pseudo open-loop phase, providing an estimation of the open loop phase of the wavefront at the pupil plane due to atmospheric turbulence. Usually the pseudo-open loop phases are estimated using wavefront sensor data and controller state (deformable mirror updates, gain, and interaction matrix). Specifically, by summing the measured wavefront sensor phase and the previous deformable mirror updates, an estimation of atmospheric phase can be made. We make make use of an alternative method using the HODM for which we have the full voltages applied to the mirror. Spatially, the HODM has comparable sampling to the SHWFS (41-by-41 actuators versus 40-by-40 sub-apertures) and therefore using the HODM for {estimating the open-loop phase} does not restrict the spatial bandwidth. Similarly, the temporal bandwidth of the system is determined by SHWFS frame rate, while the temporal bandwidth of the HODM is much higher. By using the HODM we do not lose any spatial or temporal bandwidth.}

 {From Fig.~\ref{fig:all_psd}, we can see that the SAXO provides a good correction for our data sets, as expected, therefore the HODM surface is representative of the full atmospheric phase. {We assume the residual errors are negligible owing to the high expected Strehl ratio for our conditions (80-90\% in the H band, see~\cite{Fusco_2016}). From Fig.~\ref{fig:all_psd}, we also see that the closed-loop PSDs estimated from the wavefront sensor residuals are relatively flat, indicating a good correction. We can then take the HODM surface as representative of the full atmospheric phase.} In an open-loop system, the full atmospheric phase is measured by the wavefront sensor and then used to determine the deformable mirror commands. The surface of the deformable mirror therefore represents the estimated
open-loop phase of the wavefront in the pupil plane.  It can be expressed in the form of a  lifted vector as}
\begin{equation}
    y(t)=DM_{surface}(t)
    \label{eq:pol}
,\end{equation}

 {where $DM_{surface}(t)$ is a $41^2\times1$ vector. }

 {Therefore, using a posteria data, since we know the full voltage on the DM surface we can estimate the pseudo open-loop phase.}

\end{document}